\documentclass[twoside,11pt]{article}
\usepackage{jair, theapa, rawfonts} 
\usepackage{amsmath}
\usepackage{graphicx}
\usepackage{algorithm}
\usepackage{amssymb}
\usepackage[hyphens]{url} 
\usepackage{musicography}
\usepackage{float}
\usepackage{natbib} 
\usepackage{subcaption}
\usepackage{color}
\usepackage{hyperref}
\hypersetup{
    colorlinks=true,
    linkcolor=blue,
    filecolor=blue,      
    urlcolor=blue,
    citecolor=blue
    }
\usepackage[symbol]{footmisc}

\begin{document}

\title{BacHMMachine: An Interpretable and Scalable Model for Algorithmic Harmonization}

\author {
    \name Yunyao Zhu\textsuperscript{*} \email yunyao.zhu@duke.edu \\
    \name Stephen Hahn\textsuperscript{*} \email stephen.hahn@duke.edu \\
    \name Simon Mak \email sm769@duke.edu \\
    \name Yue Jiang \email yue.jiang@duke.edu \\
    \name Cynthia Rudin \email cynthia@cs.duke.edu \\
}

\maketitle
\footnotetext{\textsuperscript{*}Equal contribution}

\begin{abstract}

Algorithmic harmonization -- the automated harmonization of a musical piece given its melodic line -- is a challenging problem that has garnered much interest from both music theorists and computer scientists. Methods for algorithmic chorale harmonization typically adopt a black-box, ``data-driven'' approach: they do not explicitly integrate principles from music theory but rely on a complex learning model trained with a large amount of chorale data. We propose instead a new harmonization model, called BacHMMachine, which employs a ``theory-driven'' framework guided by music composition principles, along with a ``data-driven'' model for learning compositional features within this framework. These music principles are elicited from the four-part Baroque chorales of J.S. Bach, but are broadly applicable for a wide range of musical genres. BacHMMachine uses a novel Hidden Markov Model based on key and chord transitions, providing a probabilistic framework for learning key modulations and chordal progressions from a given melodic line. This allows for the generation of creative, yet musically coherent harmonizations. Furthermore, integrating compositional principles allows for a much simpler model that results in vast decreases in computational burden and greater interpretability compared to state-of-the-art algorithmic harmonization methods, at no penalty to quality of harmonization or musicality. We demonstrate the improvement of BacHMMachine over existing harmonization methods via comprehensive experiments and Turing tests for Baroque chorales, and furthermore demonstrate its flexibility and applicability to other domain fields by applying our methodology to rock music. 


\end{abstract}

\noindent Harmonization is the task of generating a musically appropriate harmony given a melody as the input. While harmonization is typically viewed as a creative endeavor, it also follows formal rules set by musical convention and style. Take, for example, the harmonization of Baroque-style chorales in four voices (soprano, alto, tenor, and bass), where the given melody is a pre-existing soprano voice, and the harmony arises from the interaction of the four melodic lines. For such chorales, \textit{Gradus ad Parnassum} \citep{Fux} is an early treatise on species counterpoint, formalizing a set of musical principles for composition in multiple voices. Subject to such principles, composers may write coherent harmonizations in any number of ways, the choice of which often reflecting a composer's creative signature. 

In recent years, algorithmic harmonization has garnered notable interest from both music theorists and computer scientists. Much of the existing literature adopts a ``data-driven'' approach: musical pieces are first converted to training data, then used within state-of-the-art machine learning models for harmonization. Such methods are typically ``black-box,'' ignoring underlying compositional principles from music theory (e.g., the DeepBach model of \citealt{Hadjeres2017DeepBachAS} or the BachBot model of \citealt{Liang2017AutomaticSC}). We provide a more comprehensive review of existing harmonization models in Section \ref{sec:exist}.

Existing methods have important limitations. Because such methods use complex models for learning chorale features, they typically do \textit{not} embed the guiding principles that underlie composition. The expectation is that such models will learn these rules from data, but this learning is not perfect, and results often lack the musical coherence present in human-written harmonizations. This is particularly salient for Baroque chorales due to their highly organized harmonic structure, but is also pertinent for other musical genres. Furthermore, by ignoring structure provided by compositional principles, such harmonization models require a large amount of training data to learn this structure, not to mention idiosyncratic characteristics of given composers. For certain use cases, the training sample size needed for satisfactory model training may not even be available, resulting in unsatisfactory performance. Even when such data are available, training the harmonization model with large datasets can be computationally expensive, error-prone, and difficult to troubleshoot or tune. 


In our work, we adopt a ``theory-driven'' learning model which emulates the process an expert musician may use for harmonization \citep{harmony}. For Baroque chorales, this harmonization process typically involves (i) generating the tonal and harmonic progressions from the given melodic line and (ii) using the progressions to generate voice-leading for the remaining voices. To mimic Step (i), our proposed model learns the relationship between the given melody and local tonalities (or keys) of the chorale as well as the relationship between the melody and the underlying harmonic progression. The tonal and harmonic progressions can then be efficiently inferred via either Viterbi decoding \citep{viterbi1967error} or posterior decoding \citep{russell2002artificial}. Once the ``backbone'' of the chorale -- its tonal and harmonic progressions -- is generated, we mimic Step (ii) by building a probabilistic model for harmonization under the inferred progression subject to Baroque compositional guidelines.


Our model, which we call \textit{BacHMMachine}, provides an efficient, theory-guided, interpretable, flexible, and easy-to-tune approach to the chorale harmonization problem. Human experiments suggest preference for BacHHMachine compared to the best black-box harmonization methods. The proposed method also yields great computational savings compared to the state-of-the-art -- $\sim$30x faster than Google's Coconet \citep{Coconet_web_application} in generating chorales and $\sim$1,000x faster than the approach of \citet{Allan&Williams}. Furthermore, Turing tests suggest a surprisingly good ability to generate convincing chorales. Finally, as we can interpret the model directly, we are able to gain insights into music composition that cannot be obtained using existing methods.

Given that chorales come from the same origins as many other genres of Western music, our harmonization model naturally extends to such genres as well. In particular, we show in Section \ref{sec:rock} that the proposed BacHMMachine approach extends quite naturally to the harmonization of modern rock music. We also note that, compared to many existing methods, because our transition matrices are interpretable, humans can manually adjust them to their own taste. This would not be possible with a complex black-box approach.



\section{Musical Background\label{sec:background}}
\begin{figure}[H]
    \centering
    \includegraphics[width=\linewidth]{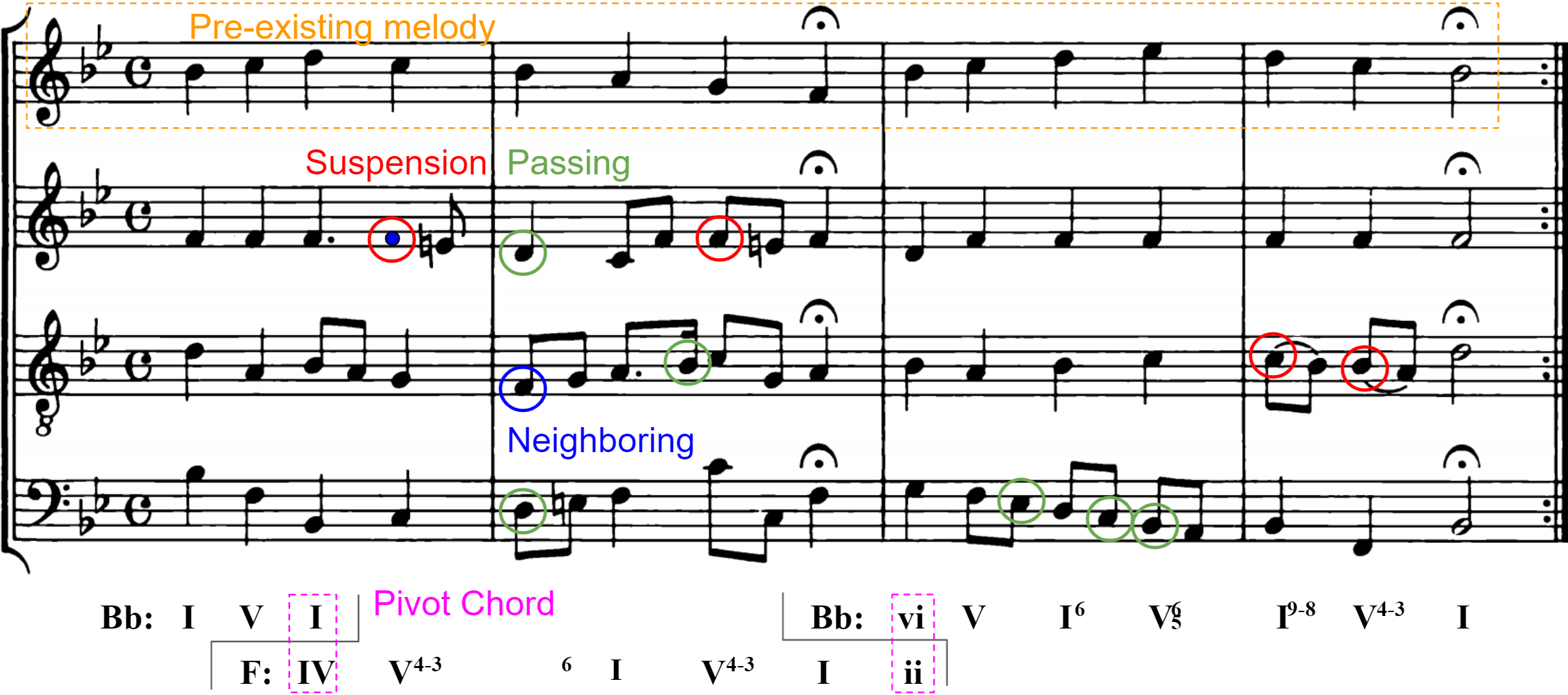}
    \caption{Annotated excerpt of J.S. Bach's chorale harmonization, \textit{Jesu, deine tiefen Wunden}, BWV 194/6}
    \label{fig:bach_chorale_analysis}
\end{figure}

\noindent We first provide background on the underlying musical structure for BacHMMachine using the four-part Baroque chorales of J.S. Bach, then show later how similar music theories can be used for harmonizing melodies of different musical genres. Figure \ref{fig:bach_chorale_analysis} shows an excerpt of one such Bach chorale. Here, the soprano (top) voice line in a chorale is the \textit{melody} of the chorale, often taken from pre-existing hymn tunes. This melody can be interpreted as the ``horizontal" aspect of a piece of music. On the other hand, \textit{harmony} involves the relationship between the notes of the chorale sung simultaneously in different voices. In this sense, harmony is often viewed as the ``vertical" aspect of music. Music theorists \citep[see, e.g.,][]{Rameau} typically classify vertical harmony using Roman numeral notation (e.g., I, IV, V), with different numerals describing unique structures. For instance, ``I'' represents triadic harmony built on the tonic scale degree (the first note of the scale), while ``V\textsuperscript{7}'' represents a ``seventh'' chord built on the dominant scale degree (the fifth note of the scale). An example is provided in Figure \ref{fig:bach_chorale_analysis}.  

\begin{figure}[H]
    \centering
    \includegraphics[width=0.75\textwidth]{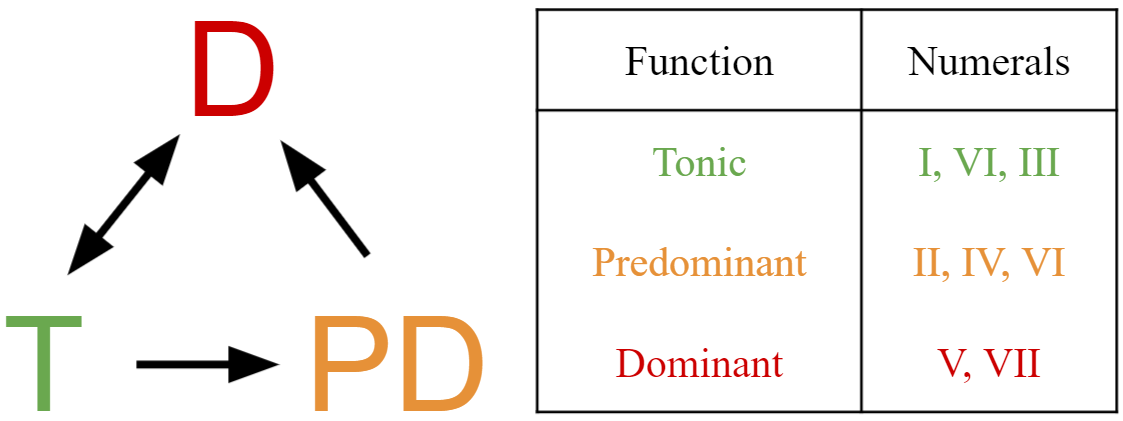}
    \caption{Visualizing the phrase model harmonic framework \citep{laitz_2016}. Here, chords (denoted by numerals) may be major, minor, or diminished.}
    \label{fig:phrase_model}
\end{figure}

Music of the Baroque era usually follows a harmonic framework known as the ``phrase model'' \citep[][see Figure \ref{fig:phrase_model}]{laitz_2016,white2018}. This framework dictates that harmonies should progress from those functioning as \textit{tonic} (I, VI, III) to those functioning as \textit{predominant} (II, IV, VI) or \textit{dominant} (V, VII). Predominants lead to dominants, and dominants resolve back to tonics. Within each functional category (tonic, predominant, or dominant), harmonies tend to progress by root relations of a descending third or fifth. Harmonic progressions that go against the phrase model are known as \textit{retrogressive} and rarely occur in Baroque music. Chorales that satisfy the phrase model yield musically cohesive harmonic progressions; those that do not may sound improperly resolved, and are musically displeasing. 

Another important aspect of chorale composition is the relationship between local tonalities. \textit{Tonality} refers to a musical passage's centricity around a single tone, where other tones in the musical environment are hierarchically related to the central pitch. Temporary transitions from one tonality (or key) to another, or \textit{modulations}, are widely used to provide a more engaging and complex harmonic structure. Like surface-level harmonic progressions, deeper tonal shifts in music also usually follow the phrase model. The sequence of tonal shifts in a chorale, or its \textit{key progression}, is integral to a chorale's musical signature. Figure \ref{fig:bach_chorale_analysis} shows the modulation from the tonic key of B-flat major to F major, then back to B-flat major.

Additional rules should be followed for pleasing and convincing chorale harmonizations. For instance, there are rules regarding chord inversions in the bass voice, and principles prescribing which chord notes can be doubled or omitted. To ensure a full sound, the soprano and alto voices and the alto and tenor voices should stay within an octave of each other. Notes in an upper voice should be relatively stable with sparing use of melodic leaps. Parallel fifths and octaves are strongly discouraged as they erode voice independence and are distracting to the listener. Non-chord tones (such as suspensions, passing, and neighboring tones) may be added to smooth voice leading and add rhythmic diversity (see Figure \ref{fig:bach_chorale_analysis}).



\section{Existing Harmonization Methods}
\label{sec:exist}



\begin{figure}[H]
\centering
\includegraphics[width=\linewidth]{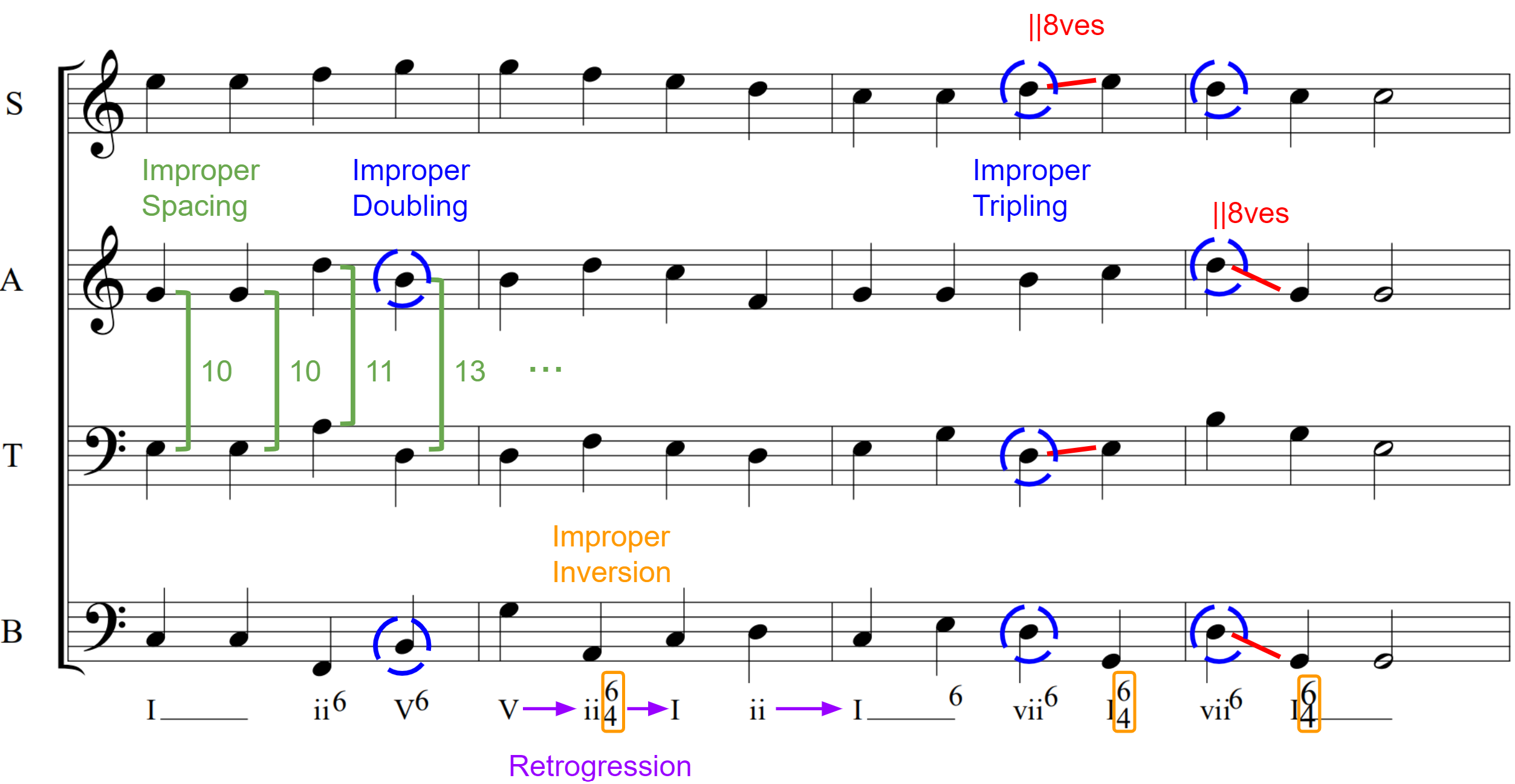}
\caption{Violations of compositional principles for the chorale harmonization in \citet{Yi2007AutomaticGO}.}
\label{FigViolations}
\end{figure}

\noindent Existing methods can be broadly grouped into those relying on Markovian models and those based on deep learning. \citet{Yi2007AutomaticGO} use a factored Markov decision processes planner to generate chordal progressions based on input melody. However, their resulting harmonization (see Figure \ref{FigViolations}) has serious flaws from a musical perspective. There are parallel octaves, unusual note doublings, dissonant leaps in the bass, awkward voice spacing and chord inversions (e.g., ending on $I^6_4$), and retrogressive harmonic progressions, which result in musically displeasing harmonies.

\citet{fixed_intermediate_chord_constraints} introduce a hierarchical modeling approach using a hidden Markov model (HMM), with user-specified fixed-chords as ``anchors,'' and generation of chords connecting them. Using a second HMM, they produce the bass voice given the chordal progression produced by the first model. Although introducing chord constraints is musically interesting, the reliance on human experts to manually insert fixed chords can make the harmonization process less flexible. Additionally, their study also fails to incorporate non-chord tones and does not always adhere to Baroque composition principles.


\citet{Allan&Williams} propose a more musically agnostic approach that uses HMMs to generate chordal progressions. A first HMM is used to generate chord notes and a second HMM adds non-chord notes decorating the harmonization. One drawback is its overly large search space, requiring over 2,800 hidden states for chorale representation! One reason is that, for this model, hidden states represent the unique sequence of intervals from the bass note. As such, there are many musically redundant hidden states that represent the same chord with different transpositions. In contrast, we adopt a more musically-informed (and much smaller) state space which captures tonality and chordal structure.

Some notable studies in chorale harmonization use deep learning methods. HARMONET \citep{Hild1991HARMONETAN} is a hierarchical architecture containing five neural nets determining chordal progression, chord inversions, the bass note, and non-chord notes. Two recent approaches using neural networks are DeepBach and BachBot. DeepBach \citep{Hadjeres2017DeepBachAS} is a graphical model that generates four-part chorales using recurrent neural networks, while BachBot \citep{Liang2017AutomaticSC} is an automatic composition system that uses a deep long short-term memory model. Both approaches are supposedly agnostic as they ``rely on little musical knowledge." Likely because of this, generated harmonizations from such approaches often violate important composition guidelines. Furthermore, these deep neural networks require a large training set and are less interpretable and harder to tune than HMMs.



\section{BacHMMachine}


What distinguishes our approach from existing work is the emphasis on emulating harmonization procedures of trained musicians that incorporate genre-specific composition conventions.
Our model includes three important novel elements: key modulation, data representation of each chord by its Roman numeral (including inversions), and strictly following chord inversion and voice-crossing constraints.

\textbf{Modulation:} 
Existing methods generally focus only on chord transitions, with chords assumed to belong to only one key throughout the entire piece.  However, chord transitions also depend on their location with respect to the whole piece and presence of modulation, and thus existing methods do not capture these musically important nuances, overlooking the prevalence and importance of key modulation in Bach chorales and other musical genres. To produce an authentic-sounding harmonization, it is crucial to take into account the key transitions in addition to the chord transitions. 

\textbf{Chord-Equivalence Representation:}
Many existing methods treat any concurrent vertical combination of notes as a unique chord. However, functionally identical chords might occur in different forms (e.g., inversions, transpositions). To reduce dimensionality of the emission and transition matrices, we treat these different forms as one, as is typical in chorale analysis. We consider only 31 unique chords in our training set and transpose all chorales to C major or A minor, greatly reducing search space and training time. We emphasize the underlying structure (chordal progression and key modulation) rather than absolute pitches of notes.
 
 
\textbf{Harmonization Constraints:} 
 For chorales, we impose basic harmonization constraints when specifying each voice line. On a chord-level, we strictly follow the chord inversion and voice crossing constraints, with combinations of notes violating them not considered in the final harmonization. On a chorale-level, once we generate possible harmonizations for the input melody, we check them for potential violations of composition guidelines and assign a penalty whenever a violation (e.g., voice crossing between chords, large jumps, parallel fifths/octaves) occurs. We output the harmonization with the lowest overall penalty. As mentioned earlier, many such harmonization principles are not exclusive to only Bach chorales, but for a wide range of musical genres (we will demonstrate this flexibility later for the harmonization of a rock song).

\subsection{Key-Chord Hidden Markov Model}
The first step in our harmonization framework is to infer plausible key and chordal progressions given an input melody. This is achieved by a novel Key-Chord HMM model which integrates tonality and chordal structure to achieve efficient, scalable and interpretable harmonization.



Let $\mathbf{M} = (m_1, \cdots, m_n)$ be the given melody line, with $m_t$ the melody note at time $t$ (i.e., on the $t^{\textrm{th}}$ beat). This can be seen as a sequence of visible states for the HMM. Let $\mathbf{K} = (k_1, \cdots, k_n)$ be the hidden key progression capturing the modulation of the chorale, with $k_t \in \mathcal{K}$ the key at time $t$, where $\mathcal{K}$ is the state space of 24 keys (12 major, 12 minor). Let $\mathbf{C} = (c_1, \cdots, c_n)$ be its hidden chordal progression, with $c_t \in \mathcal{C}$ the particular chord at time $t$, where $\mathcal{C}$ is the state space of 31 chords. We aim to recover (or \textit{decode}) the sequences of keys and chords $\mathbf{K}$ and $\mathbf{C}$ from the melody $\mathbf{M}$.

We build the Key-Chord HMM in two stages, first for the key progression, then for the chordal progression. For the key sequence $\mathbf{K}$, we impose the following first-order Markovian model on transition probabilities:
\begin{align}
\begin{split}
&\mathbb{P}( k_{t+1}|k_1, ..., k_t) = \mathbb{P}(k_{t+1}|k_t) =: T^K_{k_t,k_{t+1}},
\end{split}
\label{eq:markovkey}
\end{align}
for $t = 1, \cdots, n$. Here, $T^K_{k_t,k_{t+1}}:=\mathbb{P}(k_{t+1}|k_t)$ denotes the transition probability from key $k_t$ to $k_{t+1}$, which we estimate using chorale data. This first-order Markovian assumption is standard for HMMs, and can be justified by the earlier phrase model chordal structure.
Given key $k_t$, we assume that $m_t$, the melody note at time $t$, depends only on $k_t$, i.e.:
\begin{align}
\begin{split}
\mathbb{P}(m_t|k_1, ..., k_t, m_1, ..., m_{t-1}) = \mathbb{P}(m_t|k_t) =: E_{k_t,m_t}^K,
\end{split}
\label{eq:emissionkey}
\end{align}
for $t = 1, \cdots, n$. Here, $E_{k_t,m_t}^K$ denotes the emission probability of melody note $m_t$ from key $k_t$. This is again a standard HMM assumption, justifiable by the earlier discussion that the melody line can be well-characterized by its underlying tonality and chord quality. 

\begin{figure}[H]
\centering
\includegraphics[width=0.8\textwidth]{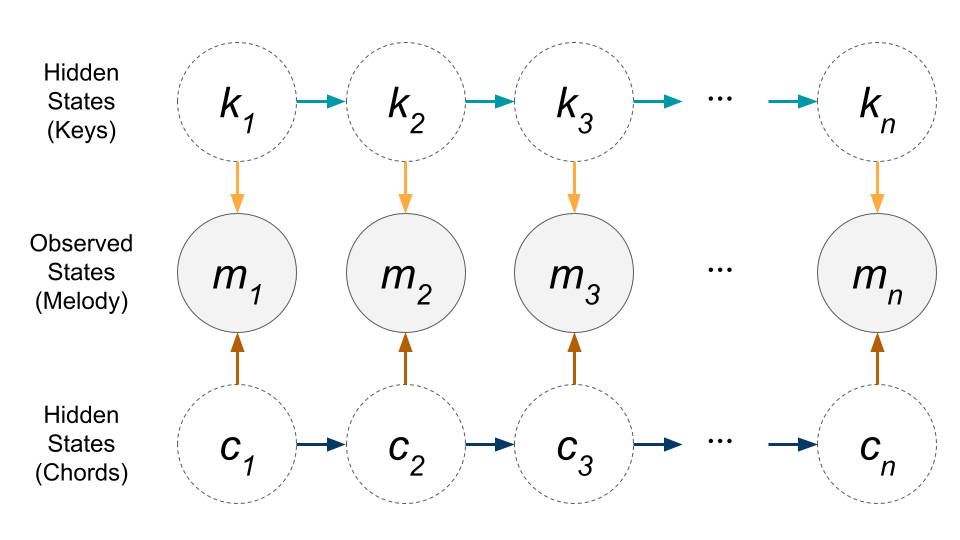}
\caption{Key-Chord HMM Visualization.}
\label{fig:keychordhmm}
\end{figure}

Next, for the chord sequence $\mathbf{C}$, we presume that the key sequence has already been decoded from data (call this inferred sequence $\mathbf{K}^*$, more on this in the next subsection). We again adopt a first-order Markovian model for transition probabilities:
\begin{align}
\begin{split}
&\mathbb{P}( c_{t+1}|c_1, ..., c_t) = \mathbb{P}(c_{t+1}|c_t) =: T^C_{c_t,c_{t+1}},
\end{split}
\label{eq:markovchord}
\end{align}
for $t = 1, \cdots, n$. Here, $T^C_{c_t,c_{t+1}}:=\mathbb{P}(c_{t+1}|c_t)$ denotes the transition probability from chord $c_t$ to $c_{t+1}$, which we estimate from data. This can again be reasoned by the earlier phrase model. Given the inferred key $k_t^*$ and chord $c_t$, we then assume that the \textit{transposed} melody note $\delta_t = m_t-k_t^*$ (i.e., modulo key change) follows the model:
\begin{align}
\begin{split}
\mathbb{P}(\delta_t|c_1, ..., c_t, \delta_1, ..., \delta_{t-1}) = \mathbb{P}(\delta_t|c_t) =: E^C_{c_t,\delta_t},
\end{split}
\label{eq:emission}
\end{align}
for $t = 1, \cdots, n$. This leverages the observation that similar harmonic structures are used over different tonalities.
Figure \ref{fig:keychordhmm} visualizes the Key-Chord HMM model: the observed states are the input soprano melody line and the hidden states are the underlying keys and chords. 


In practice, both the emission probabilities $E^K$ and $E^C$, as well as key and chord transition probabilities $T^K$ and $T^C$, must be estimated from chorale training data. We adopt the following hybrid estimation approach. First, to ensure the harmonization does not violate progressions from the phrase model, we set the probabilities of retrogressive chord transitions (i.e., those violating the phrase model) to be near zero. The remaining parameters are then estimated from the training data using maximum likelihood estimation \citep{casella2021statistical}. This ensures our model not only generates musically coherent chordal progressions in line with compositional principles, but also permits us to learn a composer's creative style under such constraints. Our proposed model requires substantially fewer parameters than existing HMM harmonization models (specifically \citealp{Allan&Williams}, which requires estimation of over $2,800^2$ transition probabilities). As shown later, this yields a computationally efficient and interpretable harmonization model, competitive with state-of-the-art models in terms of harmonization quality.



\subsubsection{Inferring Hidden Keys and Chords}

We employ two approaches for inferring the underlying hidden key-chord sequence from the Key-Chord HMM.



\textit{Viterbi Decoding}: The Viterbi decoding algorithm \citep{viterbi1967error} is a popular dynamic programming method for inferring hidden states in HMMs and is widely used in signal processing, natural language processing \citep{jurafsky2000speech}, and other fields. Here, a two-step implementation of the Viterbi algorithm allows for efficient inference of the underlying key and chord sequences.

Given melody line $\mathbf{M}$, the key inference problem can be formulated as
\begin{equation}
\mathbf{K}^* \in \text{arg}\,\max\limits_{\mathbf{K}}\,\mathbb{P}(\mathbf{K}|\mathbf{M}).
\label{eq:keyinf}
\end{equation}
Here, $\mathbb{P}(\mathbf{K}|\mathbf{M})$ is the posterior probability of a certain key sequence $\mathbf{K}$ given melody line $\mathbf{M}$ under the Key-Chord HMM. This optimization, however, involves $|\mathcal{K}|^n$ variables, which can be high-dimensional. The Viterbi algorithm provides an efficient way to solve this optimization problem via dynamic programming. In our implementation, we used the Viterbi decoding function in the Python package \texttt{hmmlearn} \citep{hmmlearn}. Similarly, given melody line $\mathbf{M}$ and and inferred key sequence $\mathbf{K}^*$, the chord inference problem can be formulated as
\begin{equation}
\mathbf{C}^* \in \text{arg}\,\max\limits_{\mathbf{C}}\,\mathbb{P}(\mathbf{C}|\mathbf{M}-\mathbf{K}^*).
\label{eq:chinf}
\end{equation}
This can again be efficiently solved via the Viterbi algorithm, with the observed states now taken to be the transposed melody $\mathbf{M}-\mathbf{K}^*$.
Algorithm \ref{alg:viterbi} outlines this two-stage Viterbi algorithm for inferring the underlying key-chord sequence $(\mathbf{K}^*,\mathbf{C}^*)$.


\begin{algorithm}[!t]
\caption{Key-Chord Viterbi decoding}
\label{alg:viterbi}
\textbf{Viterbi decoding for keys}:
\begin{itemize}
\item Set $V_0^{\rm K}(0) \leftarrow 1, V_k^{\rm K}(0) \leftarrow 0$ for all $k \in \mathcal{K}$.
\item For $t = 0, \cdots, n-1$, update for all $k \in \mathcal{K}$
\begin{align*}
V_{k}^{\rm K}(t+1) & \leftarrow \max\limits_{i \in \mathcal{K}} \left\{ V_{i}^{\rm K}(t) E^K_{k,m_{t+1}} T^K_{i,k} \right\}.
\end{align*}
\item Set $\mathbf{K}^*$ as the key sequence achieving $\max_{i \in \mathcal{K}} V_i^{\rm K}(n).$
\end{itemize}
\textbf{Viterbi decoding for chords}:
\begin{itemize}
\item Set $V_0^{\rm C}(0) \leftarrow 1, V_c^{\rm C}(0) \leftarrow 0$ for all $c \in \mathcal{C}$.
\item For $t = 0, \cdots, n-1$, update for all $c \in \mathcal{C}$
\begin{align*}
V_{c}^{\rm C}(t+1) &\leftarrow \max  \limits_{i \in \mathcal{C}}\left\{ V_{i}^{\rm C}(t) E^C_{c,m_{t+1}-k_{t+1}^*} T^C_{i,c} \right\}.
\end{align*}
\item Set $\mathbf{C}^*$ as the chord sequence achieving $\max_{i \in \mathcal{C}} V_i^{\rm C}(n).$
\end{itemize}
\end{algorithm}

\textit{Posterior Decoding}: Posterior decoding provides an alternate approach for hidden state inference. Instead of finding the key and chord sequences that maximize the \textit{joint} posterior probabilities $\mathbb{P}(\mathbf{K}|\mathbf{M})$ and $\mathbb{P}(\mathbf{C}|\mathbf{M})$, posterior encoding finds the key-chord combination $(k_t,c_t)$ that maximizes the \textit{marginal} posterior probabilities $\mathbb{P}(k_t|\mathbf{M})$ and $\mathbb{P}(c_t|\mathbf{M})$ at \textit{each} time $t$. As with the earlier key-chord Viterbi decoding, an analogous two-step procedure can be used for key-chord posterior decoding, by performing the forward-backwards algorithm (an efficient posterior decoding algorithm, see \citealt{russell2002artificial}) first for keys, then for chords. Both the Viterbi algorithm and posterior encoding are widely used for HMM decoding, and we employ both in our experiments.

\subsection{Harmonization Model}

\noindent In the second step, we use the inferred key-chord progression to generate the harmonization. From music theory, each chord admits a limited number of note arrangements over the available voices, and is subject to certain constraints to ensure a musically coherent harmonization. We embed the following constraints within our Baroque chorale model:
\begin{enumerate}
\item \textit{No voice crossings}: the harmonization must maintain the Soprano-Alto-Tenor-Bass (SATB) order vertically.
\item \textit{Voice lines should stay in their vocal range}: alto notes should be between F3-D5, tenor notes between B2-G4, and bass notes between E2-C4.
\item \textit{Voice spacing}: the soprano and alto voices should not be more than an octave apart; the same should also hold for the alto and tenor voices.
\item \textit{Chord structure}: chords should obey their specific inversion and note doubling guidelines; see \citet{harmony} for a detailed discussion of such rules.
\end{enumerate}

Given the generated progression, we enumerate \textit{chord arrangements} $\mathbf{a}_{t}^{[1]}, \mathbf{a}_{t}^{[2]}, ...$ that satisfy the above vertical constraints at time $t$. A chord arrangement $\mathbf{a}_t \in \mathbb{R}^3$ represents the alto, tenor, and bass notes. Given a selected chord arrangement $\mathbf{a}_{t}$, we iterate through all possible consecutive chord arrangement pairs $(\mathbf{a}_{t}, \mathbf{a}_{t+1}^{[1]}), (\mathbf{a}_{t}, \mathbf{a}_{t+1}^{[2]}), ...$ and select the $\mathbf{a}_{t+1}$ that minimizes the Euclidean distance between the consecutive chord arrangements. With different chord arrangements at the first beat, different harmonizations can be generated for a given chorale. In the case of a tie, we check for horizontal constraints such as voice crossings and parallel intervals. We allow occasional constraint violations (with a penalty incurred when violations are found), and select the harmonization with the lowest penalty.


\subsection{Non-Chord Notes}

\noindent We further diversify the rhythm of the generated chorale harmonization by adding \textit{non-chord notes}: ornamentation notes which do not belong to the given chordal progression. While there are a variety of non-chord tones (see \citealp{harmony}), we focus on the following types: 
\begin{enumerate}
\item \textit{Passing notes} between neighboring notes a third apart.
\item \textit{Auxiliary notes} between two repeated chord notes.
\item \textit{Appoggiaturas}, or non-chord notes on strong beats.
\end{enumerate}
These non-chord notes are added probabilistically and independently at each beat, with probabilities estimated from training chorales or specified manually.


\subsection{Data} 
\noindent
We use the Bach chorale corpus and chorale analyses provided in the \texttt{Music21} toolkit \citep{music21}, dividing chorales into those in major vs. minor keys. When processing chorale data, we treat quarter notes as the unit for one time step. Notes lasting more than one beat are split on a per-beat basis, and notes lasting less than one beat are grouped into a list of notes that together span one beat. At each time step, melody notes are represented by their corresponding MIDI pitches, chords are represented using Roman numeral notation, and keys are represented using letters. We note that the data used to fit BacHMMachine is different from the data used in past studies. Instead of using raw note data corresponding directly to pitches, we use the annotated chorales (containing chordal and modulation information, see Figure \ref{fig:bach_chorale_analysis}) which are proofread by music theorists \citep{chorale_analyses}.

For our rock data, we use the ``Corpus Study of Rock Music'' in  \citet{temperley_declercq}, which consists of analyses for the top 200 ``Greatest Songs of All Time,'' as judged by Rolling Stone magazine. \citet{temperley_declercq} provide onset timings for every melodic note and Roman numeral, as well as measure onsets. To keep the rock data consistent with the Bach chorale data, we reduce the dataset to songs that are in major keys and quadruple time (such as \musFig{4 4} or \musFig{12  8}). Keys and scale degrees are represented as integers from 0 to 11, describing absolute chromatic pitch ($C=0, C\sharp/D\flat=1,\dots, B=11$). For our study, Roman numerals were simplified to triadic, root-position form in order to reduce model complexity. We collect one representative Roman numeral, key, and melodic scale degree for every measure. The numeral, key, and scale degree of the longest duration in each measure serve as our HMM's hidden and observed states respectively. 




\section{Experiments}
In the following section we discuss our experiments regarding both the Bach-style chorale and rock harmonizations. Our experiments show how our method for chorale harmonization compares to others in terms of audience preference and training/harmonization time. We also show how BacHMMachine may be simply adjusted to work for other genres by generating harmonizations for rock melodies. All audio generated by BacHMMachine for our experiments may be found in the playlist linked below.

\[\href{https://soundcloud.com/stephen-hahn-533617325/sets/bachmmachine-harmonizations/s-jropckP2g23?si=d79e6d1be7dc4ee3b53440ac13f62868&utm_source=clipboard&utm_medium=text&utm_campaign=social_sharing}{\raisebox{-0.4\height}{\includegraphics[width=.1\linewidth]{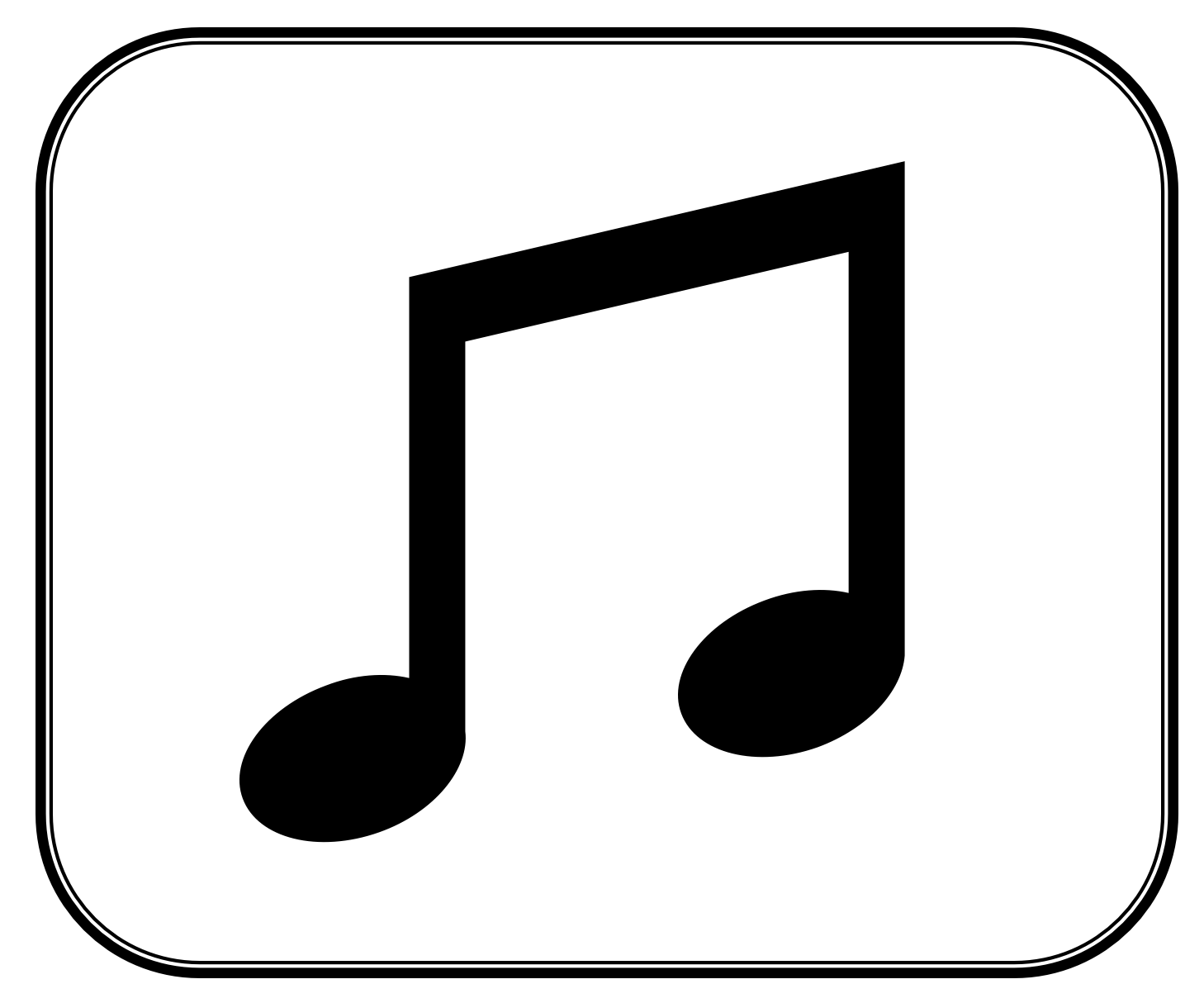}} \textrm{\;\;A playlist for audio samples of our experiments may be accessed here}.}\]

\subsection{Bach Chorale Harmonization: Set-Up}

To assess performance of BacHMMachine's chorale harmonizations, we conducted audience-preference experiments comparing our harmonizations to those generated by the existing state-of-the-art algorithmic harmonization models. The first experiment (\emph{Exp. 1}) compares the proposed BacHMMachine with the approach in \citet{Allan&Williams}, and the second experiment (\emph{Exp. 2}) compares BacHMMachine with  Google’s Coconet model \citep{huang2017counterpoint,Coconet_web_application}, which was featured on Google's homepage on Bach's 334th birthday. We further conducted a preference experiment and Turing test using Bach’s original harmonizations (\emph{Exp. 3}). Finally, we compared the BacHMMachine with either Viterbi or posterior decoding (\emph{Exp. 4}), and with or without the use of non-chord notes (\emph{Exp. 5}). Each participant evaluated chorales generated from each of these five experiments, and also one set of ``sanity-check'' questions which compare the original Bach chorale to one deliberately composed to be unpleasant and dissonant.

For each experiment, we randomly selected an existing major key chorale and used the soprano voice as the input melody. We displayed two video files, each showing the harmonization generated by one of the methods. Video files presented a scrolling score and highlighted notes as they were sung in SATB voices. We blinded the label of each harmonization. To account for order effects, the harmonization generated using BacHMMachine had a 50\% chance of being first or second. Respondents could pause and replay videos an unlimited number of times. Following the videos, we asked respondents their preference regarding the two harmonizations presented. \textit{Exp. 3} contained an additional Turing test question asking which harmonization was composed by Bach. Responses were evaluated on a five-point Likert scale that accounted for the random ordering of each harmonization pair presented.

We launched all surveys on Amazon Mechanical Turk (MTurk), requesting responses from MTurk crowd-workers with master qualifications and approval rates of \textgreater95\%.  We rejected any responses that failed the ``sanity-check'' (i.e., not indicating preference for Bach’s original harmonization over our deliberately dissonant composition). Our final analysis dataset consisted of 116 participants out of 160. Among our participants, 101 (87\%) reported enjoyment of classical music or status as music students or professionals; only 15 (13\%) reported seldom listening to classical music.

\subsection{Preference and Computational Results}

Table \ref{tab:1} shows the preference results for \textit{Exp. 1}, \textit{Exp. 2}, and \textit{Exp. 3}. We see that more respondents preferred BacHMMachine’s harmonizations to those generated by both the Allan and Williams’ implementation \citep{Allan_Williams_Python_inplementation} and Google’s Coconet application \citep{Coconet_web_application}, suggesting that the quality of BachHMMachine’s harmonization is as good as, if not better, than the existing harmonization approaches. We also see that more respondents preferred Bach’s original harmonizations to ours, which is not at all surprising. However, only 13.8\% of respondents identified Bach’s original harmonizations with certitude according to the results of the Turing test (Table \ref{tab:2}), with 36 (31\%) respondents believing to some degree that our harmonization was more likely to have been composed by Bach. This result suggests that, while BacHMMachine might not have captured all of Bach’s creative idiosyncrasies, it can produce musically convincing four-part chorale harmonizations on par with (if not better than) those generated by more complex methods. For methodological choices regarding BacHMMachine evaluated in \textit{Exp. 4} and \textit{Exp. 5}, Table \ref{tab:3} suggests that more respondents preferred our harmonizations generated using posterior decoding and with non-chord notes.

\begin{table}[t]
\centering
\begin{tabular}{lrr}
\textit{\textbf{Exp. 1: Comparison to Allan and Williams}}    & $n$ & (\%) \\ 
\hspace{0.25 cm}Definitely prefer BacHMMachine & 32 & (27.6\%)\\
\hspace{0.25 cm}Somewhat prefer BacHMMachine & 33 & (28.4\%)\\
\hspace{0.25 cm}No preference & 2 & (1.7\%)\\
\hspace{0.25 cm}Somewhat prefer Allan and Williams & 25 & (21.6\%)\\
\hspace{0.25 cm}Definitely prefer Allan and Williams & 24 & (20.7\%)\\ 
\textit{\textbf{Exp. 2: Comparison to Google's Coconet}}    &  \\
\hspace{0.25 cm}Definitely prefer BacHMMachine & 26 & (22.4\%)\\
\hspace{0.25 cm}Somewhat prefer BacHMMachine & 33 &  (28.4\%)\\
\hspace{0.25 cm}No preference & 11 &  (9.5\%)\\
\hspace{0.25 cm}Somewhat prefer Coconet & 30 & (25.9\%)\\
\hspace{0.25 cm}Definitely prefer Coconet & 16 & (13.8\%)\\ 
\textit{\textbf{Exp. 3: Comparison to original Bach}}    &  \\ 
\hspace{0.25 cm}Definitely prefer BacHMMachine & 20 & (17.2\%)\\
\hspace{0.25 cm}Somewhat prefer BacHMMachine & 23 & (19.8\%)\\
\hspace{0.25 cm}No preference & 9 & (7.8\%)\\
\hspace{0.25 cm}Somewhat prefer original Bach & 41 & (35.3\%)\\
\hspace{0.25 cm}Definitely prefer original Bach & 23 & (19.8\%)\\ 
\end{tabular}
\caption{Comparison of harmonization approaches: \textit{``Which harmonization do you prefer?''}}
\label{tab:1}
\end{table}

\begin{table}[t]
\centering
\centering
\begin{tabular}{lrr}
 & $n$ & (\%)\\
\hspace{0.25 cm}Definitely BacHMMachine & 10 & (8.6\%)\\
\hspace{0.25 cm}Somewhat BacHMMachine & 26 & (22.4\%)\\
\hspace{0.25 cm}No preference & 22 & (19.0\%)\\
\hspace{0.25 cm}Somewhat original Bach & 42 & (36.2\%)\\
\hspace{0.25 cm}Definitely original Bach & 16 & (13.8\%)
\end{tabular}
\caption{Turing test: \textit{``Which was composed by Bach?''}}
\label{tab:2}
\end{table}

\begin{table}[t]
\centering
\begin{tabular}{lrr}
\textit{\textbf{Exp. 4: Posterior decoding vs. Viterbi}}    & $n$ & (\%) \\ 
\hspace{0.25 cm}Definitely prefer Posterior decoding & 24 & (20.7\%)\\
\hspace{0.25 cm}Somewhat prefer Posterior decoding & 33 & (28.4\%)\\
\hspace{0.25 cm}No preference & 16 & (13.8\%)\\
\hspace{0.25 cm}Somewhat prefer Viterbi & 26 & (22.4\%)\\
\hspace{0.25 cm}Definitely prefer Viterbi & 17 & (14.7\%)\\ 
\textit{\textbf{Exp. 5: Inclusion of non-chord notes}}    &  \\
\hspace{0.25 cm}Definitely prefer \textit{no} non-chord notes & 16 & (13.8\%)\\
\hspace{0.25 cm}Somewhat prefer \textit{no} non-chord notes & 32 &  (27.6\%)\\
\hspace{0.25 cm}No preference & 15 & (12.9\%)\\
\hspace{0.25 cm}Somewhat prefer \textit{with} non-chord notes & 30 & (25.9\%)\\
\hspace{0.25 cm}Definitely prefer \textit{with} non-chord notes & 23 & (19.8\%)
\end{tabular}
\caption{Comparison of BacHMMachine decoding approaches: \textit{``Which harmonization do you prefer?''}}
\label{tab:3}
\end{table}

Table \ref{tab:4} reports the computation time (in seconds) needed to train each of the three algorithms on the same eight chorales, as performed on a personal computer using a 1.8 GHz Dual-Core Intel Core i5 processor. The reported results are the averages of 5 algorithm runs. \textit{We see that the training time for the Allan and Williams' algorithm is $\sim$1,000x that of the proposed BacHMMachine.} After both models are trained, the harmonization of a new chorale can be performed very quickly (under a second). Unfortunately, we cannot compare the model training time for Google's Coconet, since the application was already pre-trained. However, \textit{the computation time for Coconet's trained model to harmonize a new chorale is $\sim$30x longer than that of BacHMMachine}. This significant edge in computation time (either for model training or harmonization) for BacHMMachine shows that the proposed approach is indeed much more scalable than the state-of-the-art. This is again due, in large part, to the integration of compositional principles within the harmonization model, which allows for model reduction and efficient training.

\begin{table}[H]
\centering
\begin{tabular}{lrr}
\textbf{\textit{Approach}} & Train time & Harmonization time\\
\hspace{0.25 cm}BacHMMachine & 6.27 & 0.60\\
\hspace{0.25 cm}Allan and Williams & 6175.05 & 0.56\\
\hspace{0.25 cm}Google's Coconet & (unknown) & 18.56
\end{tabular}
\caption{Comparison of training time on 8 chorales and harmonization time (given the trained model) in seconds.}
\label{tab:4}
\end{table}

\subsection{Model Interpretability}

\begin{figure}[t]
\centering
\includegraphics[width=\textwidth]{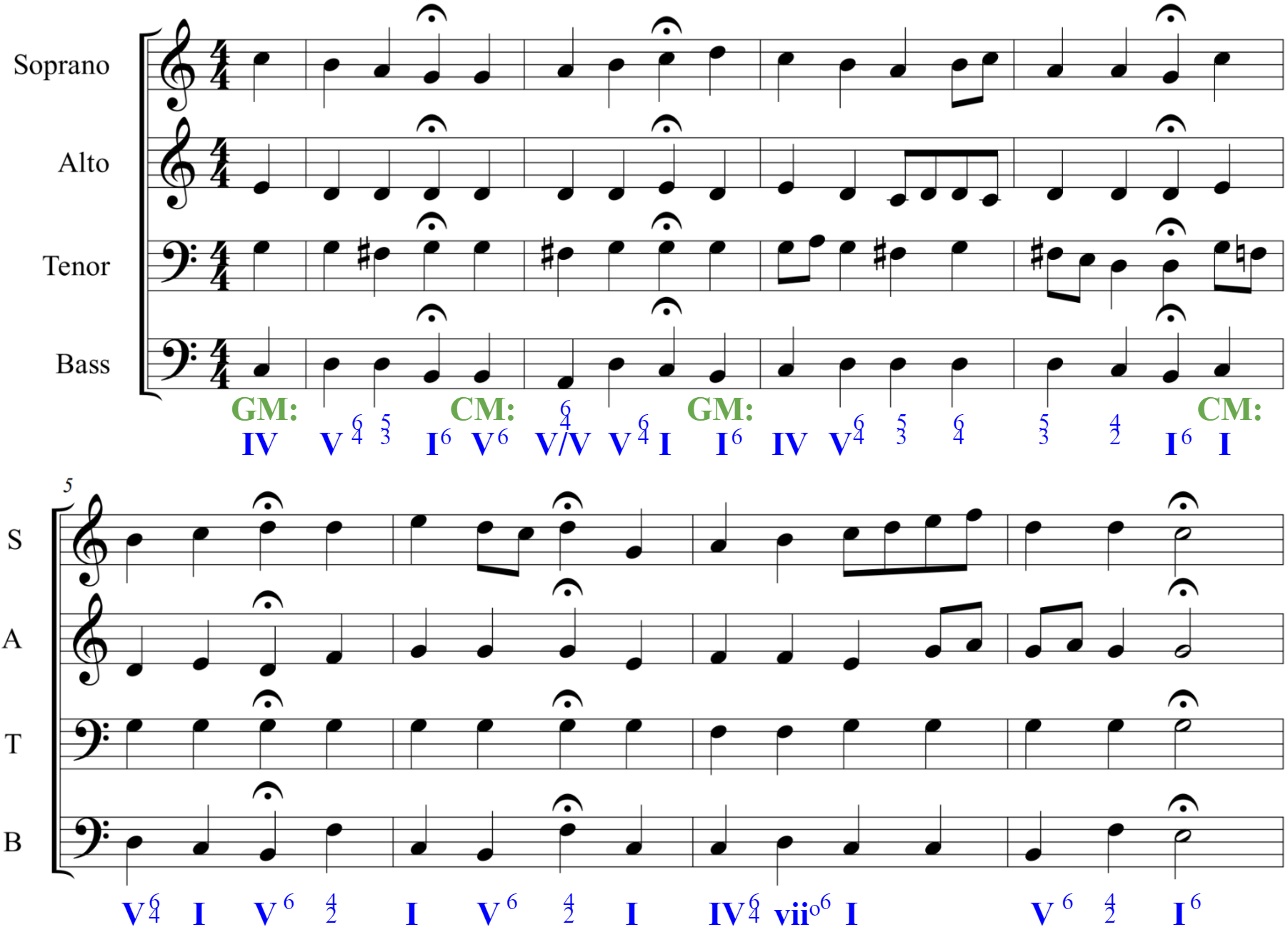}
\caption{Annotated excerpt of the BacHMMachine harmonization of melody from J.S. Bach's chorale \emph{Ach Gott und Herr}, BWV 255. 
\label{fig:harmonizpost}
}

\end{figure}

Figure \ref{fig:harmonizpost} shows an excerpt of a harmonized chorale using BacHMMachine, with posterior decoding and including non-chord notes. A careful musical analysis of this shows the generated harmonization satisfies much of the compositional principles desired for Baroque chorale composition. We investigate this further below.

\begin{figure}[t]
    \centering
    \includegraphics[width = 0.5\textwidth]{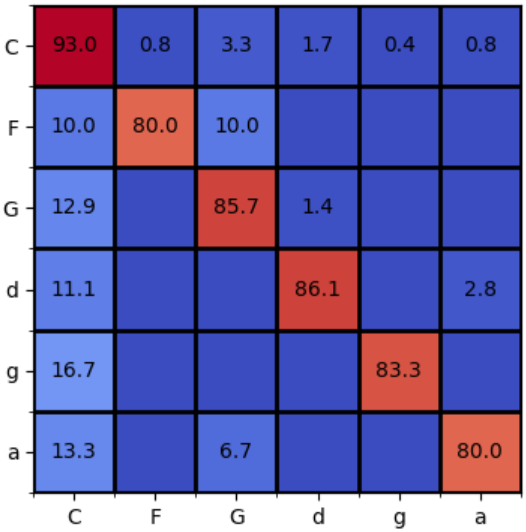}
    \caption{Estimated key transition probabilities (\%) in the trained BacHMMachine model.}
    \label{fig:key_transitions}
\end{figure}

Figure \ref{fig:key_transitions} visualizes the key transition probabilities learned by BacHMMachine. There are several observations of interest. First, as expected, we see that the estimated key probabilities strongly conform to the phrase model. Second, we see that key transition probabilities demonstrate a high degree of key stability, meaning that the generated chorales only change keys approximately once per phrase. This frequency of key changes is similar to Bach chorale harmonizations. Finally, we also observe that all keys have a strong probability of modulating back to the tonic, with the next most common destination being the dominant, which again is in line with musical expectation \citep{laitz_2016}. This shows that, by integrating compositional principles within the harmonization model, BacHMMachine can learn interpretable stylistic features which are verifiable from music theory. 

\[\href{https://soundcloud.com/stephen-hahn-533617325/bachmmachine-harmonizations-bwv255-posterior-decoding/s-dmKUueTAJ3r?in=stephen-hahn-533617325/sets/bachmmachine-harmonizations/s-jropckP2g23&utm_source=clipboard&utm_medium=text&utm_campaign=social_sharing}{\raisebox{-0.4\height}{\includegraphics[width=.1\linewidth]{images/MusicButton1.png}} \textrm{\;\;Listen to BacHMMachine's harmonization of the melody, \textit{Ach Gott und Herr}}.}\]

\begin{figure}[t]
    \centering
    \begin{subfigure}[b]{0.7\textwidth}
         \includegraphics[width=\linewidth]{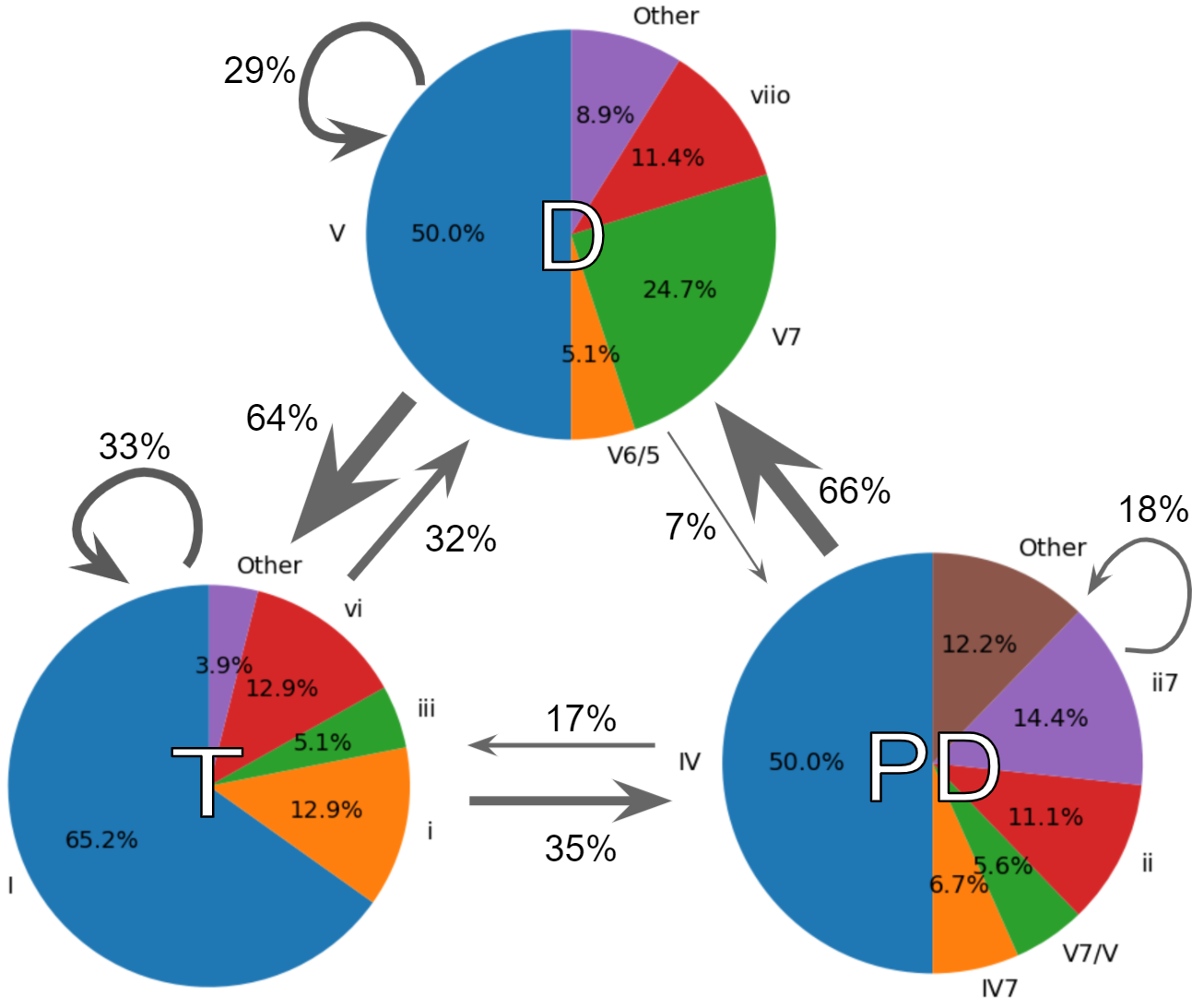}
        \subcaption{Estimated chord transition probabilities in the BacHMMachine model trained on Bach chorale harmonizations.}
        \label{fig:functional_transitions}
     \end{subfigure}
     \hfill
     \begin{subfigure}[b]{0.7\textwidth}
         \includegraphics[width=\linewidth]{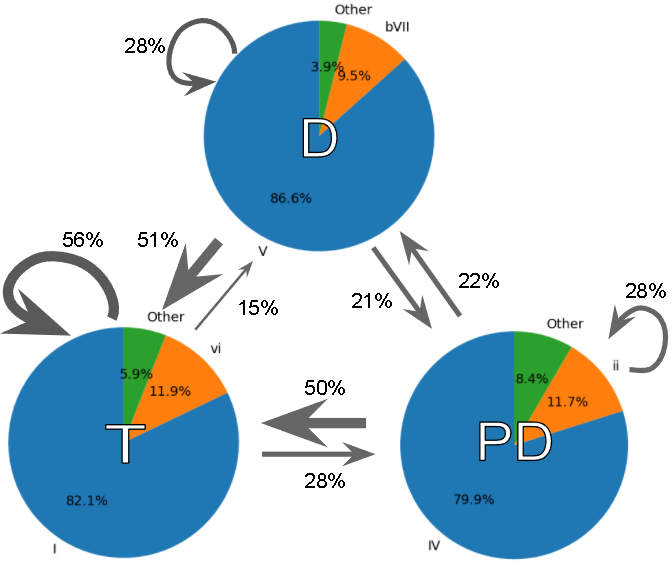}
        \subcaption{Estimated chord transition probabilities in the BacHMMachine model trained on \citet{temperley_declercq}'s rock analyses within \citet{laitz_2016}'s phrase model.}
        \label{fig:functional_transitions_rock}
     \end{subfigure}
    \caption{Estimated chord transition probabilities in the BacHMMachine model, grouped by tonic (T), predominant (PD) and dominant (D) chords.}
    \label{fig:bachmmachine_functional_transitions}
\end{figure}

Figure \ref{fig:functional_transitions} visualizes the chord transition probabilities learned by BacHMMachine, organized by the tonic (T), predominant (PD) and dominant (D) chord groups from the phrase model (see Figure \ref{fig:phrase_model}). We see that, as expected, the estimated chord probabilities from BacHMMachine closely follow the harmonic structure dictated by the phrase model. In particular, the model learned that tonics progress to tonics, predominants, or dominants with near equal probability, that most predominants progress to dominants, and that most dominants progress to tonics. This agrees with expected chord transitions from the phrase model \citep{laitz_2016}, which demonstrates the interpretability of the proposed model in corroborating stylistic features of Baroque chorales.

\subsection{Harmonization of Rock Melodies}
\label{sec:rock}

The BacHMMachine model is not only capable of harmonizing Bach chorales, but also a wide range of musical genres that follow similar harmonization principles. To demonstrate this flexibility, we apply this model for harmonizing rock songs. 

Our procedure for harmonizing rock melodies is similar to that for Bach chorales. First, we generate the underlying key progression and transpose the melody according to key. Then we determine the harmonic progression of the melody (one numeral for each user-defined measure). The harmonic and key progressions determine the bass line, which plays the harmonic root on every downbeat then arpeggiates the other chord tones until the measure is full. Electric piano plays appropriate block chords or arpeggiates the harmony. Finally, we add a simple drum pattern to fill out the texture. Each instrument plays a common rock rhythm, which may be determined stochastically from a rhythmic vocabulary, or manually selected by the user.

The audio file below contains the BacHMMachine harmonization of the melody from Procol Harum's \textit{A Whiter Shade of Pale}. From a quick listen, we find that the generated harmonization provides a convincing imitation of a typical rock harmonic progression. The syntactical harmonic structure of rock music is still a hot topic for debate in music theory, so there is no consensus on a generalized rock ``phrase model'' (We observe how the rock transitions relates to the Baroque phrase model in Figure \ref{fig:bachmmachine_functional_transitions}). However, it is clear that the blues, which often follows a straightforward harmonic formula, had a heavy influence on the development of rock (as blues artist Muddy Waters said, ``the blues had a baby and named it rock and roll'' \citep{muddywaters}). Specifically, the blues is based on variations of a simple twelve-bar progression: 

$$
\vert\vert :\quad I\quad\vert\quad I\quad\vert\quad I\quad\vert\quad I\quad\vert\vert\quad
IV\quad\vert\quad IV\quad\vert\quad I\quad\vert\quad I\quad\vert\vert\quad
V\quad\vert\quad IV\quad\vert\quad I\quad\vert\quad I\quad:\vert\vert
$$

\noindent The trained BacHMMachine model determined that the most probable outcome for melody harmonization consists of simple progressions involving ``I,'' ``IV,'' and ``V'' in a single key (see Figure \ref{fig:functional_transitions_rock}). Hence, the progressions learned by BacHMMachine sound very similar to that from a blues progression, and can thus successfully imitate rock harmony.

\subsection{Interpretability Permits User Interactions}
A benefit of BacHMMachine's model interpretability is the fact that humans can edit the transition probabilities if desired. In fact, the rock progressions primarily involving ``I,'' ``IV,'' and ``V'' in a single key might be undesirable for some users. The simple structure of BacHMMachine allows us to manipulate the transition matrices to produce more interesting and personal results. Our playlist includes a purely data-driven harmonization and another harmonization after manually adjusting the transition probabilities to involve a greater chord diversity. 

\[\href{https://soundcloud.com/stephen-hahn-533617325/bachmmachine-harmonization-a-whiter-shade-of-pale/s-HltCil56dBU?si=a3e6de762a194e5da642b9ea7a342e82&utm_source=clipboard&utm_medium=text&utm_campaign=social_sharing}{\raisebox{-.40\height}{\includegraphics[width=.1\linewidth]{images/MusicButton1.png}} \textrm{\;\;Purely data-driven harmonization}.}\]

\[\href{https://soundcloud.com/stephen-hahn-533617325/bachmmachine-harmonization-a-whiter-shade-of-pale-v2/s-VPVTowY0CBE?in=stephen-hahn-533617325/sets/bachmmachine-harmonizations/s-jropckP2g23&si=3139f23ef59e44668fa2fc72e5905832&utm_source=clipboard&utm_medium=text&utm_campaign=social_sharing}{\raisebox{-.40\height}{\includegraphics[width=.1\linewidth]{images/MusicButton1.png}} \textrm{\;\;Harmonization with manually adjusted Roman numeral transition matrix}.}\]

\noindent The custom matrix simply boosted the probability of encountering non-major diatonic harmonies (ii, iii, vi, vii\textsuperscript{o}) as well as $\flat$VII, all of which were largely avoided by the data-driven transition matrix. This kind of simple adjustment to the transition matrix can be executed quickly and provides tremendous potential for diverse melody harmonizations.

\section{Conclusion}\label{sec:concl}

In this study, we described a probabilistic framework capable of generating musically convincing harmonizations. The main strength of the model is that it is musically informed, emulating the harmonization process of a human composer by incorporating musical guidelines and constraints. Because of this, we are able to reduce the number of violations of composition guidelines. By using professional analyses instead of raw musical pitches as input data, our method requires considerably fewer hidden states and takes a tiny fraction of the training time of other HMM approaches. The use of HMMs themselves leads to the generation process being more interpretable, with faster composition times compared to recent deep learning approaches.

Because we directly use musical information, it is straightforward to extend our model to take advantage of additional considerations. For instance, we may explore supplemental data-encoding schemes that discretize chorales into smaller time units, add an additional layer regarding key modulation or pivot chords, or consider higher-order Markovian models. Regardless, it is clear that by incorporating musical principles in our model, we are able to achieve high-quality harmonization with much simpler and easily-interpretable models at a tiny fraction of the computational cost. We encourage future researchers to consider such domain-specific information when designing generative models to hopefully achieve similar results.

\newpage

\bibliographystyle{plainnat}
\bibliography{Paper.bib}

\begin{thebibliography}{23}
\providecommand{\natexlab}[1]{#1}
\providecommand{\url}[1]{\texttt{#1}}
\expandafter\ifx\csname urlstyle\endcsname\relax
  \providecommand{\doi}[1]{doi: #1}\else
  \providecommand{\doi}{doi: \begingroup \urlstyle{rm}\Url}\fi

\bibitem[Allan and Williams(2004)]{Allan&Williams}
Moray Allan and Christopher K.~I. Williams.
\newblock Harmonising chorales by probabilistic inference.
\newblock In \emph{Proceedings of Neural Information Processing Systems}, 2004.

\bibitem[Andrews and Sclater(1993)]{harmony}
William~G. Andrews and Molly Sclater.
\newblock \emph{Materials of Western Music: Part 1}.
\newblock Gordon V. Thompson Music, 1993.

\bibitem[Casella and Berger(2021)]{casella2021statistical}
George Casella and Roger~L Berger.
\newblock \emph{Statistical Inference}.
\newblock Cengage Learning, 2021.

\bibitem[Cuthbert and Ariza(2021)]{music21}
Michael~Scott Cuthbert and Christopher Ariza.
\newblock Music21: A toolkit for computer-aided musicology and symbolic music
  data, 2021.
\newblock URL \url{https://github.com/cuthbertLab/music21}.

\bibitem[Dinculescu et~al.(2019)Dinculescu, Huang, Cooijmans, Roberts,
  Courville, and Eck]{Coconet_web_application}
Monica Dinculescu, Cheng-Zhi~Anna Huang, Tim Cooijmans, Adam Roberts, Aaron
  Courville, and Douglas Eck.
\newblock Coconet coucou, 2019.
\newblock URL \url{http://coconet.glitch.me/}.

\bibitem[Fux(1725)]{Fux}
Johann~Joseph Fux.
\newblock \emph{Gradus ad Parnassum (Steps or Ascent to Mount Parnassus)}.
\newblock W. W. Norton \& Company, 1725.

\bibitem[Hadjeres et~al.(2017)Hadjeres, Pachet, and
  Nielsen]{Hadjeres2017DeepBachAS}
Ga{\"e}tan Hadjeres, F.~Pachet, and F.~Nielsen.
\newblock Deep{B}ach: a steerable model for {B}ach chorales generation.
\newblock In \emph{Proceedings of the 34th International Conference on Machine
  Learning}, 2017.

\bibitem[Hild et~al.(1991)Hild, Feulner, and Menzel]{Hild1991HARMONETAN}
H.~Hild, J.~Feulner, and W.~Menzel.
\newblock {HARMONET}: A neural net for harmonizing chorales in the style of
  {J}. {S}. {B}ach.
\newblock In \emph{Proceedings of Neural Information Processing Systems}, 1991.

\bibitem[Huang et~al.(2017)Huang, Cooijmans, Roberts, Courville, and
  Eck]{huang2017counterpoint}
Cheng-Zhi~Anna Huang, Tim Cooijmans, Adam Roberts, Aaron Courville, and Douglas
  Eck.
\newblock Counterpoint by convolution.
\newblock In \emph{International Society for Music Information Retrieval
  ({ISMIR})}, 2017.

\bibitem[Jones et~al.(2021)Jones, Tymoczko, and Robb]{chorale_analyses}
Andrew Jones, Dmitri Tymoczko, and Hamish Robb.
\newblock Music21 corpus: Bach chorale analyses, 2021.
\newblock URL
  \url{https://github.com/cuthbertLab/music21/tree/master/music21/corpus/bach/choraleAnalyses}.

\bibitem[Jurafsky(2000)]{jurafsky2000speech}
Dan Jurafsky.
\newblock \emph{Speech \& Language Processing}.
\newblock Pearson Education India, 2000.

\bibitem[Kaliakatsos-Papakostas and
  Cambouropoulos(2014)]{fixed_intermediate_chord_constraints}
Maximos~A. Kaliakatsos-Papakostas and E.~Cambouropoulos.
\newblock Probabilistic harmonization with fixed intermediate chord
  constraints.
\newblock In \emph{Proceeding of the Joint 11th Sound and Music Computing
  Conference (SMC) and 40th International Computer Music Conference (ICMC)},
  2014.

\bibitem[Laitz(2016)]{laitz_2016}
Steven Laitz.
\newblock \emph{The Complete Musician: An Integrated Approach to Tonal Theory,
  Analysis, and Listening}.
\newblock Oxford University Press, 2016.

\bibitem[Lebedev et~al.(2021)Lebedev, Lee, Varoquaux, and Farrow]{hmmlearn}
Sergei Lebedev, Anthony Lee, Gael Varoquaux, and Chris Farrow.
\newblock hmmlearn: Unsupervised learning and inference of {H}idden {M}arkov
  {M}odels, 2021.
\newblock URL \url{https://github.com/hmmlearn/hmmlearn}.

\bibitem[Liang et~al.(2017)Liang, Gotham, Johnson, and
  Shotton]{Liang2017AutomaticSC}
Feynman~T. Liang, Mark Gotham, Matthew Johnson, and J.~Shotton.
\newblock Automatic stylistic composition of {B}ach chorales with deep {LSTM}.
\newblock In \emph{International Society for Music Information Retrieval},
  2017.

\bibitem[Mammana et~al.(2019)Mammana, Nisoli, Moray, and
  Williams]{Allan_Williams_Python_inplementation}
Lorenzo Mammana, Eric Nisoli, Allan Moray, and Christopher Williams.
\newblock Generation of {B}ach chorales harmonisation using {H}idden {M}arkov
  {M}odels, 2019.
\newblock URL \url{https://github.com/lorenzomammana/py-bach-harmonisation}.

\bibitem[Rameau(1722)]{Rameau}
Jean-Philippe Rameau.
\newblock \emph{Treatise on Harmony}.
\newblock Dover Publications, 1722.

\bibitem[Russell and Norvig(2002)]{russell2002artificial}
Stuart Russell and Peter Norvig.
\newblock \emph{Artificial Intelligence: A Modern Approach}.
\newblock Pearson, 2002.

\bibitem[Temperley and deClercq(2011)]{temperley_declercq}
David Temperley and Trevor deClercq.
\newblock A corpus analysis of rock harmony.
\newblock \emph{Popular Music}, 30:\penalty0 47--70, 2011.
\newblock URL \url{http://rockcorpus.midside.com/index.html}.

\bibitem[Viterbi(1967)]{viterbi1967error}
Andrew Viterbi.
\newblock Error bounds for convolutional codes and an asymptotically optimum
  decoding algorithm.
\newblock \emph{IEEE Transactions on Information Theory}, 13\penalty0
  (2):\penalty0 260--269, 1967.

\bibitem[Waters(1977)]{muddywaters}
Muddy Waters.
\newblock The blues had a baby and they named it rock and roll, pt. 2, 1977.

\bibitem[White and Quinn(2018)]{white2018}
Christopher~WM White and Ian Quinn.
\newblock Chord context and harmonic function in tonal music.
\newblock \emph{Music Theory Spectrum}, 40\penalty0 (2):\penalty0 314--335, 11
  2018.
\newblock ISSN 0195-6167.
\newblock \doi{10.1093/mts/mty021}.
\newblock URL \url{https://doi.org/10.1093/mts/mty021}.

\bibitem[Yi and Goldsmith(2007)]{Yi2007AutomaticGO}
Liangrong Yi and J.~Goldsmith.
\newblock Automatic generation of four-part harmony.
\newblock In \emph{Proceedings of the Conference on Uncertainty in Artificial
  Intelligence}, 2007.

\end{thebibliography}
\end{document}